\documentclass[aps,preprint,showpacs,preprintnumbers,amsmath,amssymb,floatfix]{revtex4}
\usepackage{graphicx}
\begin{document}
\def\brho {\mbox{\boldmath $\rho$}}
\def\r {{\bf r}}

\title{Role of excited states in the splitting 
dynamics of interacting Bose-Einstein condensates when ramping-up a barrier}

\author{Alexej I. Streltsov, Ofir E. Alon and Lorenz S. Cederbaum}

\affiliation{Theoretische Chemie, Physikalisch-Chemisches Institut, Universit\"at Heidelberg,\\
Im Neuenheimer Feld 229, D-69120 Heidelberg, Germany}

\begin{abstract}
An essentially-exact approach to compute the wavefunction in the time-dependent 
many-boson Schr\"odinger equation is derived and
employed to study accurately the process of splitting a trapped condensate
when ramping-up a barrier such that a double-well is formed.
We follow the role played by many-body excited states during the splitting process.
Among others, a 'counter-intuitive' regime is found in which 
the evolution of the condensate when the barrier is ramped-up 
sufficiently slow {\it is not} to the ground-state which is a fragmented condensate, 
but to a low-lying excited-state which is a coherent condensate.
Experimental implications are discussed.
\end{abstract}

\maketitle

The first realizations of Bose-Einstein condensate in ultra-cold dilute gases have boosted the community
to explore matter-wave phenomena and how to manipulate and utilize them. 
An ultimate goal of researchers is to be able to design, 
realize and detect a desired quantum state of the many-atom system.
One of the most popular 'screenplays' studied is splitting of Bose-Einstein condensates (BECs) 
by deforming a single well to a double-well, 
see \cite{e1,e2,e3,e35,e4} for experimental and \cite{t0,t1,t2,t3,t4,t5} 
for theoretical works.  
In such scenarios, 
the many-atom system is often prepared in the ground-state of a harmonic trap
and a central barrier is ramped-up to a certain fixed height.
As the trap transforms from harmonic to double-well geometry the time-dependent state  
of the many-boson system continuously changes its localization from the center of the initial trap 
to two separated parts localized around the minima of the double-well.
Side by side, it can also change its character -- from condensed to a 
two-fold fragmented state \cite{t2}. 
Attacking the splitting process, 
much attention has been paid to understanding when it is adiabatic \cite{t0,t1,t2},
demonstrating that the slower the barrier is ramped-up (to a certain fixed height),
the closer is the BEC to the ground-state of the bosons in the double-well.

In the present work we study the dynamics of splitting 
when ramping-up a barrier beyond the presently available 
theoretical and computational approaches.
We develop and report on an essentially-exact and
numerically-efficient approach for the solution of 
the time-dependent many-boson Schr\"odinger equation, 
which we term multi-configurational time-dependent Hartree for bosons (MCTDHB).
Applying MCTDHB to the ramping-up-a-barrier problem
we follow the many-boson wavefunction throughout the
splitting process and identify the role and impact of many-body {\it excited-states} 
on the splitting process.
Among others, we identify a new 'counter-intuitive' regime where
the evolution of the condensate when the barrier in ramped-up 
sufficiently slow {\it is not} to the ground-state of the double-well
which is a fragmented BEC, 
but to a low-lying excited-state which is a coherent BEC.
More details are given below.

Our starting point is 
the many-body Hamiltonian describing $N$ interacting bosons in 
a trap, $\hat H = \sum_{k=1}^N \left[\hat T(\r_k) + V(\r_k,t)\right] +
 \sum_{k>l=1}^N U(\r_k-\r_l)$. Here, $\r_k$ is the coordinate of the $k$-th particle,
$\hat T(\r)$ and $V(\r,t)$ stand for the kinetic energy and trap potential,
respectively, and $U(\r_k-\r_l)$ describes the pairwise interaction between the $k$-th and $l$-th atoms. 
To solve the time-dependent Schr\"odinger equation
$\hat H \Psi = i \frac{\partial\Psi}{\partial t}$ we write the many-body wavefunction $\Psi$ 
as a linear combination of {\it time-dependent} permanents
\begin{equation}\label{Ansatz}
 \Psi(\r_1,\r_2,\ldots,\r_N,t) = \sum_{\vec{n}} C_{\vec{n}}(t) \Phi_{\vec{n}}(\r_1,\r_2,\ldots,\r_N,t). \
\end{equation}
In the representation (\ref{Ansatz}),
the time-dependent permanents $\Phi_{\vec{n}}$ are constructed by distributing the $N$ bosons 
over $j=1,\ldots,M$ {\it time-dependent} one-particle functions (orbitals) $\{\phi_j(\r,t)\}$,
and the summation runs over all possible occupations $\vec{n}$ preserving 
the total number of bosons $N$.

To proceed, we utilize the Dirac-Frenkel variational principle \cite{DF}
and after some lengthy but straightforward algebra obtain a set of coupled non-linear, 
generally integro-differential equations 
for the coefficients $C_{\vec{n}}(t)$ and orbitals $\left\{\phi_j(\r,t)\right\}$,
for all $\vec{n}$ and $j=1,\ldots,M$, 
\begin{eqnarray}\label{MCTDH_eq}
 & & \hat{\mathbf P} \left[ \left\{\hat T(\r) + V(\r,t) \right\}
 \phi_j(\r,t) + \sum_{qksl} 
\left\{\brho(t)\right\}^{-1}_{jq} \rho_{qksl}(t)
     U_{kl}(\r,t) \phi_s(\r,t) \right] =  i \frac{\partial\phi_j(\r,t)}{\partial t}, \nonumber \\
 & & \qquad \qquad \sum_{\vec{n}'} \left< \Phi_{\vec{n}}\left|\hat H\right|\Phi_{\vec{n}'}\right> C_{\vec{n}'}(t)
 =  i \frac{d C_{\vec{n}}(t)}{dt}.  \
\end{eqnarray}
The quantities $\brho(t)=\left\{\rho_{qs}(t)\right\}$ 
and $\rho_{qksl}(t)$ appearing in (\ref{MCTDH_eq}) 
are the matrix elements of the reduced one- and two-body densities of $\Psi$, 
$\rho(\r_1|\r_1';t)=\sum^M_{qs} \rho_{qs}(t) \phi^\ast_q(\r'_1,t)\phi_s(\r_1,t)$ and
$\rho(\r_1\r_2|\r'_1,\r'_2;t)=\sum^M_{qksl} \rho_{qksl}(t)
\phi^\ast_q(\r'_1,t) \phi^\ast_k(\r'_2,t) \phi_s(\r_1,t) \phi_l(\r_2,t)$, respectively. 
The local time-dependent potentials $U_{kl}(\r,t)=\int \phi_k^\ast(\r',t)U(\r-\r')\phi_l(\r',t)d\r'$
originate from the two-body interaction. 
Finally, the operator $\hat{\mathbf P}=1-\sum_{k=1}^M\left|\phi_k(\r,t)\left>\right<\phi_k(\r,t)\right|$
appearing on the left-hand-side of Eq.~(\ref{MCTDH_eq}) is a projection operator 
which ensures that the orbitals remain orthogonal to one another throughout 
the propagation in time \cite{MCTDHB}.

The MCTDHB theory gathered in Eqs.~(\ref{Ansatz},\ref{MCTDH_eq}) 
is capable by construction to describe 
the evolution of $N$-bosons' wavefunctions and thus many-body properties of BECs in time-dependent potentials, 
and is an essentially-exact time-dependent theory. 
We stress that it derives from a full, time-dependent variational principle 
where no restrictions except of orthonormality 
on the coefficients $C_{\vec{n}}(t)$ and orbitals 
$\left\{\phi_j(\r,t)\right\}$ are employed.
The key idea behind expansion (\ref{Ansatz}) is that, 
by allowing {\it both} the expansion coefficients 
$C_{\vec{n}}(t)$ and orbitals $\{\phi_j(\r,t)\}$ used to construct the permanents 
to be {\it time-dependent}, a significant computational advantage in solving the 
many-body Schr\"odinger equation arises in comparison to 
many-body expansions with orbitals of fixed shape, also see below.
It is gratifying to mention that the present many-body propagation theory 
adapts to identical bosons the multi-configurational time-dependent 
Hartree approach routinely used for multi-dimensional dynamical systems 
consisting of distinguishable particles \cite{MCTDH123}.
By explicitly exploiting bosons' statistics it is possible
to successfully and quantitatively attack the dynamics of 
a large number of bosons with the present theory \cite{MCTDHB}.
Finally, we can also propagate the MCTDHB equations (\ref{MCTDH_eq}) in imaginary
time and compute for static (time-independent) traps self-consistent ground and 
excited eigenstates of bosonic systems,
since in that case (\ref{MCTDH_eq}) boil down to 
the general variational many-body theory 
for interacting bosons developed recently in \cite{MCHB}.

Let us study now the splitting dynamics of a repulsive bosonic system
from an initial harmonic trap where the ground eigenstate 
is a slightly depleted condensed state, to the final double-well potential where the 
ground-state many-body wavefunction is (totally) two-fold fragmented.
Specifically, we consider $N=200$ $^{\mathrm 87}$Rb atoms initially
prepared in an elongated, quasi-one-dimensional harmonic trap of longitudinal 
$\omega_{\|}=2\pi \times 44.7$Hz and transverse $\omega_{\perp}=2\pi \times 1.1$kHz frequencies.
At time $t=0$ a barrier of Gaussian shape is ramped-up linearly in time
to a height of $V_{max}$ and with ramp-up time of $T_{ramp}$. 
Introducing a convenient length scale of $L=1{\mathrm {\mu m}}$,
we translate to dimensionless units in which
the kinetic energy reads
$\hat T(x) = -\frac{1}{2}\frac{\partial^2}{\partial x^2}$,
the time-dependent trap potential is
$V(x,t)=\frac{x^2}{2\sigma^2} + V_{max}\exp\left(-\frac{x^2}{2\sigma^2}\right)\times
\left\{t/T_{ramp}, \ t \le T_{ramp}; 1, \ t > T_{ramp}\right\}$,
and the effective atom-atom interaction is $U(x-x')=\lambda_0\delta(x-x')$ 
where the transverse confinement is properly accounted for \cite{Max}.
The values of the parameters are $\sigma=2.6$ and $V_{max}=30$, 
corresponding to an inter-well separation of $13.6{\mathrm \mu m}$ at the end of the ramping-up process,
and $\lambda_0=0.1$. 
Finally, time is expressed in units of $\frac{mL^2}{\hbar}=1.37$msec, 
where $m$ is the mass of $^{\mathrm 87}$Rb atom, 
and energy in units of $\frac{\hbar^2}{mL^2}=116$Hz.

\begin{figure}
\includegraphics[width=8.6cm, angle=-0]{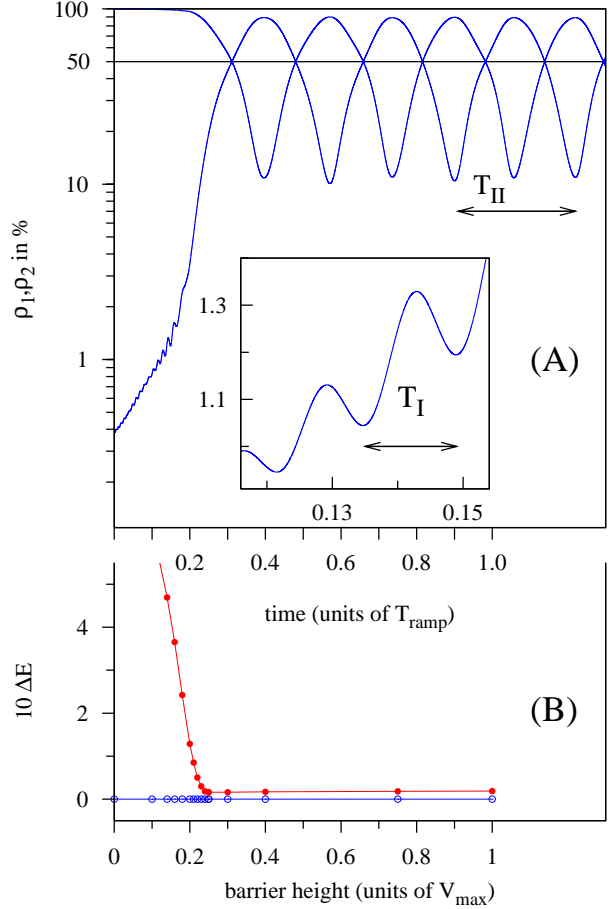}
\caption {(Color online) Ramping-up a barrier for $N=200$ bosons ($^{\mathrm 87}$Rb atoms) 
with $\lambda_0=0.1$ and ramping-up time $T_{ramp}=1000$.
During the ramping-up process the many-body wavefunction evolves 
from a condensed towards two-fold fragmented state.
(A) Natural occupation numbers $\rho_1(t)$ and $\rho_2(t)$ evaluated 
at each time-point are plotted on a logarithmic scale. 
Inset shows oscillatory behavior of the occupation numbers around $t_I=0.14T_{ramp}$.
Periods of the oscillations are attributed to local excitation times, 
$T_{local}=\frac{2 \pi}{\Delta E}$,
see text for more details.
(B) Energy gap $\Delta E$ between the many-body ground- and first excited-state 
computed as a function of barrier height corresponding to 
different time-points of the time-dependent trap potential $V(x,t)$. 
The energy of the ground state is taken as reference energy. 
Time is expressed in units of $\frac{mL^2}{\hbar}=1.37$msec
and energy in units of $\frac{\hbar^2}{mL^2}=166$Hz.
}
\label{fig1}
\end{figure}

We begin our investigations by ramping-up the barrier during a time of $T_{ramp}=1000$ 
until the final double-well is achieved.
The developed MCTDHB method is used to solve the time-dependent many-body Schr\"odinger equation
during and following the ramping-up process. 
Since visualization of the time-dependent many-body wavefunction is quite cumbersome,
we prescribe its natural occupation numbers,
i.e., eigenvalues of the corresponding reduced one-particle density 
$\rho(x|x';t)=\sum^M_j \rho_{j}(t)\phi^{\ast{NO}}_j(x',t)\phi^{NO}_j(x,t)$,
at each point in time \cite{t2,MCHB}.  
We have found that for the range of parameters considered here,
MCTDHB with two orbitals accurately describes
the time-dependent many-body splitting dynamics. 
The corresponding natural occupation numbers $\rho_1(t)$ and $\rho_2(t)$
are plotted in Fig.~\ref{fig1}A as a function of time.
The initial state is a slightly depleted condensed state ($0.38\%$) 
because at $t=0$ only one natural orbital is macroscopically occupied.
As the barrier is ramped-up,
$\rho_2(t)$ increases with time, i.e., the many-body wavefunction 
becomes more and more depleted.
At approximately one third of the ramping-up time $\rho_1(t)=\rho_2(t)$,
indicating that the system is momentarily two-fold fragmented.
From now on and although the barrier is ramped-up further, 
the occupation numbers oscillate around this totally fragmented configuration, 
as has been obtained in \cite{t2}. 
It is worth noticing, however, that an oscillatory behavior of the occupation numbers
exists during all the splitting process with different amplitudes and 
frequencies, see Fig.~\ref{fig1}A.

To get a deeper insight into the physics of these oscillations 
we compute via imaginary time propagation of MCTDHB 
several low-lying many-body eigenstates of the static system at 
different barrier heights along the ramping-up path.
The energy gap $\Delta E$ between the many-body ground- and first excited-state
is plotted in Fig.~\ref{fig1}B as a function of barrier height. 
Since the height of the barrier increases linearly with time
we can measure the evolution in the units of either ramping-up times or barrier heights.   
We show that the oscillation frequency of the natural orbitals' occupations 
can be attributed to $\Delta E$.
We define local excitation times $T_{local}=\frac{2 \pi}{\Delta E}$ and 
compare them with the corresponding oscillation periods at different barrier heights.

Here we present a comparison at two different points in time: 
at $t_I=0.14 T_{ramp}$ and $t_{II}=T_{ramp}$.
In the inset of Fig.~\ref{fig1}A we plot on an enlarged scale
the oscillations of the natural occupation numbers around $t_I$. 
The period of oscillations deduced from this inset is $T_I=13.7$. 
Since the ramping-up process is linear,
$t_I$ corresponds to the barrier height of $V_I=0.14V_{max}$. 
The corresponding energy gap $\Delta E_I=0.47$ 
gives the value of the local excitation time $T_{local}=\frac{2\pi}{\Delta E_I}=13.4$, 
which is close to the observed  oscillation period of the natural orbitals for this time-point.
Analogously, at $t_{II}$ the observed oscillation period 
of $T_{II}=317.9$ agrees with the respective local 
excitation time of $T_{local}=333.5$ obtained from the energy gap of $\Delta E_{II}=0.0188$.
We obtain reasonable agreement between corresponding 
local excitation times and oscillation periods for other times as well.
These results can readily be explained: 
the initial state is the ground-state and the ramping-up process
studied is quite slow, implying that at each time-point
the propagated state is quite close to the corresponding ground-state.
This observation also indicates an adiabatic character of the studied ramping-up process,
and that only one many-body excited-state is primarily involved in the dynamics.

Let us now analyze the amplitudes of the oscillations of the natural orbitals.
From Fig.~\ref{fig1}A we see that oscillations are very modest
at the beginning of the ramping-up process and substantial
at the end of the process (notice the logarithmic scale).
This is in accord with the decreasing energy gap between
the ground and first excited states depicted in Fig.~\ref{fig1}B.
As expected, an ideal adiabatic ramping-up along the
ground-state trajectory is favored by
a large energy gap $\Delta E$. 
It is achieved for ramping-up times $T_{ramp}$
large compared to the local excitation time $T_{local}=\frac{2 \pi}{\Delta E}$.

\begin{figure}[ht]
\includegraphics[width=8.6cm,angle=-0]{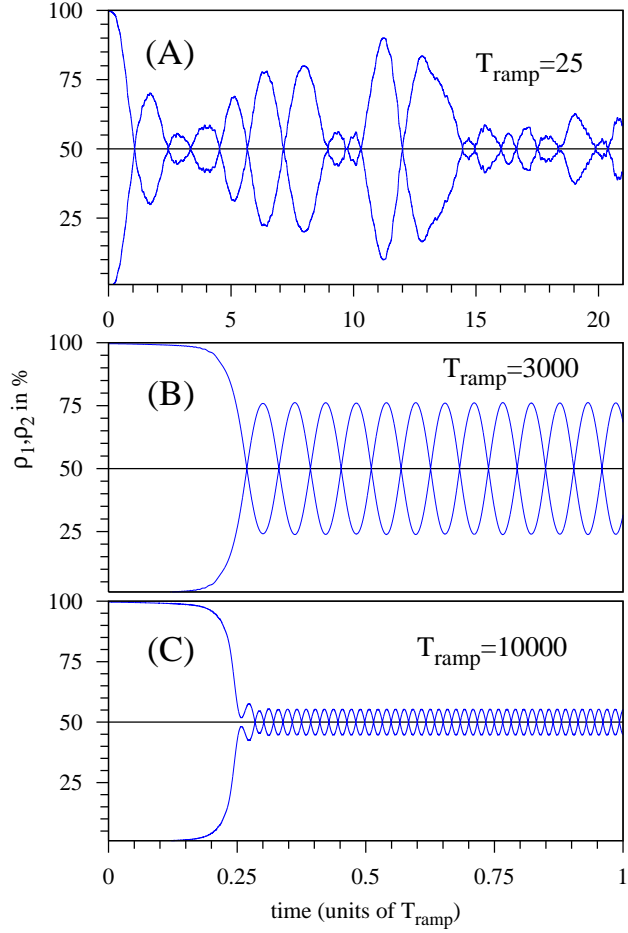}
\caption [kdv]{(Color online) 
Quantifying how difficult it is to (fully) fragment a condensate.
Plotted are the natural occupation numbers for $N=200$ bosons ($^{\mathrm 87}$Rb atoms) 
with $\lambda_0=0.1$ and for three ramping-up times $T_{ramp}$. 
Increasing $T_{ramp}$ the time-dependent many-body state approaches the ground-state of 
the double-well potential which is a (fully) fragmented BEC.
Time is expressed in units of $\frac{mL^2}{\hbar}=1.37$msec.
For $T_{ramp}=10000=13.7$sec which 
is of the order of a condensate's lifetime 
the amplitude of oscillations is still about $10\%$.
}
\label{fig2}
\end{figure}

To study quantitatively the adiabatic character of the ramping-up process,
we repeat the computation for several different durations of the ramping-up process.
The results for $T_{ramp}=25, 3000, 10000$ are presented in Fig.~\ref{fig2}.
From this figure a significant suppression of the amplitudes of the oscillations 
with growing ramping-up times is clearly seen. 
For the studied system of $N=200$ bosons,
a ramping-up procedure as long as of $T_{ramp}=10000=13.7$sec
still leads to 10.7\% fluctuations of the final state. 
We see that even for such a long ramping-up time,
which is of the order of the lifetime of a condensate \cite{e4}, 
the final state deviates noticeably from the respective eigenstate 
and the ramping-up process is still away from being ``ideal adiabatic''.

In the above example the properties of the ground state and lowest excited state change smoothly
with barrier height.
For small barrier heights these states are condensed, 
the excited state being more depleted than the ground state.
As the barrier grows,
these states become two-fold fragmented 
and the first excited state results from the ground state by a transfer
of a boson from one fragment to another.  
However, this simple scenario can vary strongly
with the number of particles and with their interaction strength \cite{MCHB}.

\begin{figure}[ht]
\includegraphics[width=8.6cm,angle=-0]{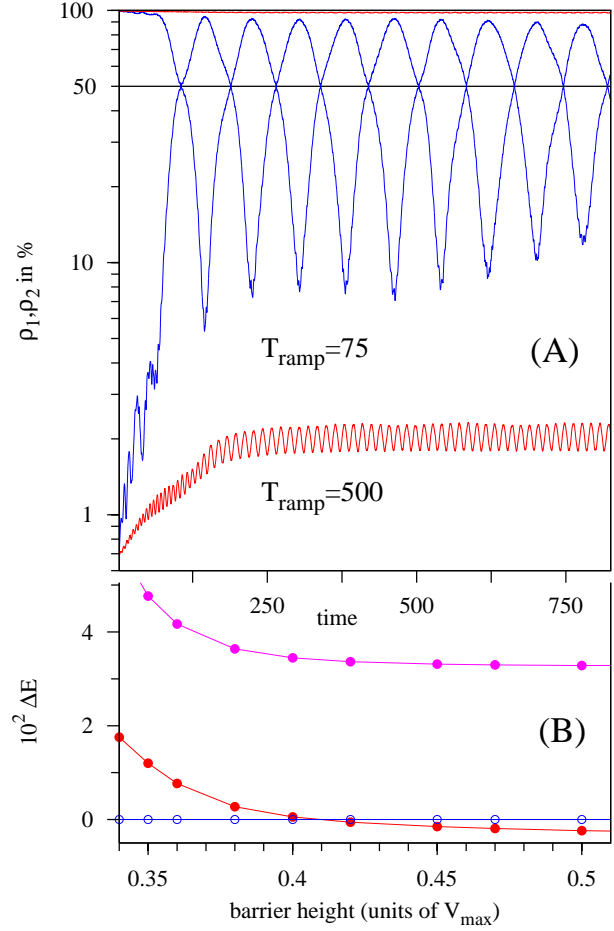}
\caption [kdv] {(Color online)
Same as in Fig.~\ref{fig1} except for the stronger interaction strength of $\lambda_0=0.3$. 
(A) For $T_{ramp}=75$ the many-body wavefunction evolves 
from a condensed towards the two-fold fragmented ground state. 
For {\it longer} durations of the ramping-up, e.g., for $T_{ramp}=500$,
a 'counter-intuitive' regime is found in which 
the evolution of the condensate {\it is not} to the fragmented ground-state , 
but to a low-lying coherent excited state.
(B) Static energy gaps $\Delta E$ as a function of barrier height.  
Around $V_{cr}\approx0.41V_{max}$ 
the condensed ground-state and lowest-excited fragmented state 
come very close to each other and interchange their order,
see text for more details.
Time is expressed in units of $\frac{mL^2}{\hbar}=1.37$msec
and energy in units of $\frac{\hbar^2}{mL^2}=166$Hz. 
}
\label{fig3}
\end{figure}

To proceed,
we consider the same experiment as before, 
but with a three-times stronger interaction strength $\lambda_0=0.3$.
This is achieved by a tighter confinement of $\omega_{\perp}=2\pi \times 3.3$kHz.
Again, the ground state of the final double-well for this system is fully two-fold fragmented.
In Fig.~\ref{fig3}A we plot $\rho_1(t)$ and $\rho_2(t)$
for the durations $T_{ramp}=75$ and $T_{ramp}=500$ of the ramping-up process.
In both cases, the initial wavefunction is the ground eigenstate in the harmonic trap
which is a slightly ($0.68\%$) depleted condensed state, see Fig.~\ref{fig3} at $t=0$.
The evolution of $\rho_1(t)$ and $\rho_2(t)$ for the faster ramping-up process of $T_{ramp}=75$
looks like the ``adiabatic'' dynamics studied in the previous example
but it is not such a dynamics.
Indeed, prolonging the ramping-up time, e.g., to $T_{ramp}=500$,
the time-dependent solution does not at all evolve towards the fragmented ground state, 
but rather to an intermediate state which, 
according to the occupation number analysis, 
remains condensed all the time.

To arrive at a deeper insight into the physics behind this 'counter-intuitive' regime,
which we call {\it inverse} regime, 
we again investigate the lowest eigenstates of the static double-wells 
at different barrier heights corresponding to different ramping-up times.
The relevant map of static energy gaps $\Delta E$ 
is plotted in Fig.~\ref{fig3}B as a function of barrier height.  
This map has very interesting features -- around 
some critical barrier height $V_{cr}\approx0.41V_{max}$ 
the ground state and lowest excited state come very close to each other and interchange their order.
Such a behavior signifying a very narrow avoided crossing (the width of which we cannot detect here)
can appear only between states of very different physical origin.
Indeed, at $V=0.40V_{max}$, the ground state is a slightly depleted 
condensed state $\rho_1=98.14\%$, $\rho_2=1.86\%$
while the first excited is almost a totally two-fold fragmented state: 
$\rho_1=51.0\%$, $\rho_2=49.0\%$.
At $V=0.42V_{max}$ we observe an inverse situation: 
the ground state is now two-fold fragmented $\rho_1=50.32\%$, $\rho_2=49.68\%$,
while the first excited state is condensed $\rho_1=98.12\%$, $\rho_2=1.88\%$.

The following picture of the ramping-up-a-barrier process emerges in the inverse regime.
For a slow ramping-up process,
only one quantum eigenstate is essentially populated although another state 
is crossing or very close by.
Clearly, because of the very different physical nature of both states,
the initially populated state cannot abruptly change its properties 
during the relevant time interval
and the presence of the partner state essentially does not influence the dynamics.
On the other hand, for a faster ramping-up process, more excited eigenstates
are involved in the evolution of the many-body wave-function and
a coupling between these eigenstates allows the system to overcome the crossing point 
and evolve towards the lowest eigenstate which is a true two-fold fragmented ground state.

Let us briefly summarize. 
We show that the dynamics of splitting of an ultra-cold bosonic cloud by ramping-up a barrier
depends on the duration of the process and on the (effective) interaction strength between the bosons.
There are (at least) two distinct regimes:
(i) an {\it adiabatic} regime where the initial condensed ground-state evolves towards 
the ground two-fold fragmented eigenstate of the final double-well potential and 
asymptotically approaches it with increasing ramping-up time;
(ii) an {\it inverse} regime where the initial condensed state 
evolves towards the ground two-fold fragmented eigenstate only for short ramping-up times, 
while for slow ramping-up processes the time-dependent state stays condensed during all the evolution
and thereby evolves to a {\it non}-ground many-body eigenstate. 
The physical insight on these regimes follows from the analysis of the
low-lying many-body excited-states taken at different times.
The above findings were made possible by developing
a multi-configurational time-dependent Hartree theory for bosons (MCTDHB) 
capable of providing a quantitative description of the time evolution of the bosonic systems. 
MCTDHB opens up further possibilities to explore the challenging 
many-body dynamics of many-boson systems.

\begin{acknowledgments}
We thank H.-D. Meyer for fruitful discussions.
Financial support by the DFG is gratefully acknowledged.
\end{acknowledgments}

\end{document}